\documentclass[aps,prl,twocolumn,showpacs]{revtex4}
\usepackage{amsmath}
\usepackage{amssymb}
\usepackage{graphicx}

\begin{document}

\title{Phase equilibrium in two orbital model under magnetic field}
\author{S. Dong}
\author{X. Y. Yao}
\author{K. F. Wang}
\author{J.-M. Liu}
\email{liujm@nju.edu.cn}
\affiliation{Nanjing National Laboratory of Microstructures, Nanjing University, Nanjing
210093,China\\
International Center for Materials Physics, Chinese Academy of
Sciences,Shenyang, China}
\date{\today}

\begin{abstract}
The phase equilibrium in manganites under magnetic field is studied using a
two orbital model, based on the equivalent chemical potential principle for
the competitive phases. We focus on the magnetic field induced melting
process of CE phase in half-doped manganites. It is predicted that the
homogenous CE phase begins to decompose into coexisting ferromagnetic phase
and CE phase once the magnetic field exceeds the threshold field. In a more
quantitative way, the volume fractions of the two competitive phases in the
phase separation regime are evaluated.
\end{abstract}

\maketitle

Manganites, a typical class of strong correlated electron systems, have been
intensively studied in the last decade, due to their unusual behaviors of
potential applications such as colossal magnetoresistance (CMR). The double
exchange (DE) mechanism can explain the magnetic transition qualitatively,
but more complex mechanism responsible for CMR is not yet fully understood.
The idea of phase separation (PS) was recently proposed to understand the
physics essence underlying the amazing behaviors of manganites, while more
and more theoretical and experimental evidences confirm the existence of PS
due to the intrinsic inhomogeneity \cite{Dago,review,Uehara}.

Former investigations on the phase diagram of manganites revealed the
first-order character of phase transitions between various phases, e.g.
charge-ordered (CO) insulator and ferromagnetic (FM) metal \cite%
{review,Loudon,Mura,Aliaga}. The insulator-metal transition in manganites
can be reasonably understood as the consequence of percolation of FM metal
filaments embedded in the insulated matrix, and there are plenty of
experimental evidences to support this PS framework \cite%
{review,Zhang,Tokunaga}. Current theories on manganites mainly stem from the
competition between several interactions: DE, super exchange, Hund coupling,
electron-phonon interaction and Coulomb interaction \cite{review}. Besides,
the effect of quench disorder on PS dynamics is highlighted, especially on
the large scale PS. The theoretical progress has enabled us to sketch the
phase diagram in some special regimes from calculation and identify the PS
regime in parameter space with various microscopic models \cite%
{Yunoki1,Yunoki2,Yunoki3}. Nevertheless, it is still unclear theoretically
how the PS develops, especially under an external perturbation, e.g.
magnetic or electric field. In other word, \emph{it is of interest to not
only identify the existence of PS regime, but also concern how the phase
separation occurs and how it evolves upon external perturbation}, because
potential applications call for more sufficient theoretical interpretation.
For instance, the CMR effect, which is one of the most attracting topics in
the physics of manganites, may be described by the resistor network model
phenomenologically, based on the percolation mechanism. In such case the
volume fraction of metal phase is the key input variable, which, however,
lacks credible theoretical investigation yet. In earlier studies, this
important variable was obtained from experiment or toy model \cite{Mayr,Dong}%
.

In this letter, we attempt to study the phase equilibrium (PE) in a two
orbital model. We emphasize particularly the evolvement of PS upon the
magnetic field perturbation. So we call it PE instead of PS in this work.
Let's begin with a simplified model Hamiltonian which has been frequently
used \cite{review}:

\begin{eqnarray}
Hami. &=&-\sum\limits_{\boldsymbol{i\alpha }\gamma \gamma ^{\prime }\sigma
}t_{\gamma \gamma ^{\prime }}^{\boldsymbol{\alpha }}d_{\boldsymbol{i}\gamma
\sigma }^{\dag }d_{\boldsymbol{i}\gamma ^{\prime }\sigma }-J_{H}\sum_{%
\boldsymbol{i}}\boldsymbol{s}_{_{\boldsymbol{i}}}\cdot \boldsymbol{S}_{i}
\nonumber \\
&&+J_{AF}\sum_{<\boldsymbol{i},\boldsymbol{j}>}\boldsymbol{S}_{i}\cdot
\boldsymbol{S}_{j}+\boldsymbol{H\cdot }\sum_{\boldsymbol{i}}g\boldsymbol{(s}%
_{i}+\boldsymbol{S}_{i})
\end{eqnarray}%
where $\boldsymbol{\alpha }$ is the vector connecting nearest-neighbor (NN)
sites and $d_{\boldsymbol{i}\gamma \sigma }^{\dag }$($d_{\boldsymbol{i}+%
\boldsymbol{\alpha }\gamma ^{\prime }\sigma }$) is the generation
(annihilation) operator for $e_{g}$ electron with spin $\sigma $ in the $%
\gamma $($\gamma ^{\prime }$)-orbital on site $\boldsymbol{i}$($\boldsymbol{i%
}+\boldsymbol{\alpha }$);\ $t_{rr^{\prime }}^{\boldsymbol{\alpha }}$ is the
NN hopping amplitude between $\gamma $ and $\gamma ^{\prime }$-orbital ($%
d_{x^{2}-y^{2}}$ as $a$ orbital, $d_{3z^{2}-r^{2}}$ as $b$ orbital) along $%
\boldsymbol{\alpha }$-direction, with $t_{bb}^{z}=t_{0}>0$ ($t_{0}$ is taken
as energy unit), $t_{aa}^{x}=t_{aa}^{y}=\frac{3}{4}t_{0}$, $%
t_{bb}^{x}=t_{bb}^{y}=\frac{1}{4}t_{0}$, $%
t_{ab}^{y}=t_{ba}^{y}=-t_{ab}^{x}=-t_{ba}^{x}=\frac{\sqrt{3}}{4}t_{0}$, $%
t_{aa}^{z}=t_{ab}^{z}=t_{ba}^{z}=0$, respectively; $\boldsymbol{S}_{i}$ is
the spin operators for $t_{2g}$ core on site $\boldsymbol{i}$, while $%
\boldsymbol{s}_{i}$ for $e_{g}$ itinerant electron. The first term
represents the kinetic energy (DE process) which leads to FM spin
arrangement. The second term is the Hund coupling of $e_{g}$ and $t_{2g}$
electrons where $J_{H}>0$ is large enough to be regarded as infinite, so the
spin of $e_{g}$ electron is always parallel with the same-site $t_{2g}$
spin. The super exchange interaction $J_{AF}>0$ prefers to couple NN $t_{2g}$
spins antiferromagnetically; The last term represents the magnetic field
contribution with magnetic field $H$ and Lande factor $g$. The
electron-phonon coupling and Coulomb repulsion are not taken into account in
this Hamiltonian and their effect on PE will be discussed below.

\begin{figure}
\includegraphics[width=0.5\textwidth,trim=15 13 10 25]{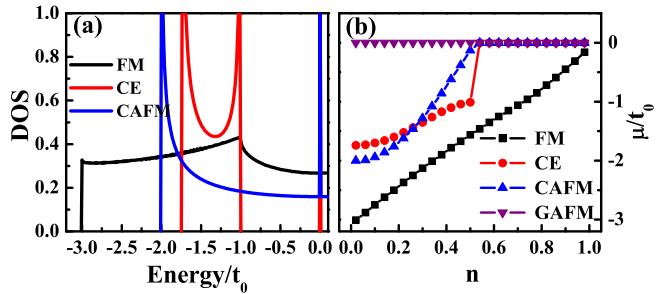}
\caption{\label{Fig.1}(color online) (a)Density of state of
FM/CE/CAFM phase. Here only regime $E\leq 0$ is displayed since the
symmetrical $E>0$ part is empty in ground state. In GAFM, there is
only a single energy level at $0$ because of the localization
of $e_{g}$ electron. (b)Chemical potential $\protect\mu $ as a function of $%
e_{g}$ electron concentration. The energy gap $t_{0}$ for the CE phase at $%
n=0.5$ is evident both in (a) and (b).}
\end{figure}

The above simplified Hamiltonian can be solved exactly once a prior $t_{2g}$
spin pattern is given. In real manganites, various $t_{2g}$ spin patterns
exist corresponding to abundant phases, e.g. FM, antiferromagnetic (AFM),
CO, orbital-ordered (OO) phases. In this work, several typical $t_{2g}$
patterns confirmed from experiments: C-type AFM (CAFM), G-type AFM (GAFM),
FM and CE phase are chosen as the candidates to compare. The CAFM phase is
constructed by antiferromagnetically coupled one-dimension FM lines, while
the CE phase is constructed by antiferromagnetically coupled one-dimension
zigzag FM chains and is found to be CO/OO \cite{Wollan,Good,Brink}. The GAFM
takes the familiar AFM arrangement in all three directions. Then the
Hamiltonian can be exactly solved when the Hund factor $J_{H}$ is simplified
as infinite. The procedure of derivation is straightforward and the details
can be found in Ref.\cite{review}. Then density of state (DOS, $D(E)$) of
the these phases can be calculated from the dispersion relationship using
numerical method, as shown in Fig.\ref{Fig.1}(a). In addition, the chemical
potential $\mu $ of these phases is obtained simultaneously by integrating
the DOS, as shown in Fig.\ref{Fig.1}(b). Consequently, the ground state
energy is calculated. For instance, the energy of FM phase can be written
as:
\begin{equation}
\frac{E(n,H)}{N}=\frac{1}{N}\int^{E<\mu }D(E)EdE+\frac{27}{2}J_{AF}-3H
\label{E}
\end{equation}%
here $N$ is the number of whole sites and is infinite ideally. The first
integral term gives the energy of all $e_{g}$ electrons; term $27J_{AF}/2$
arises from the six NN FM correlation between $|\boldsymbol{S}|=3/2$ $t_{2g}$
cores in classical approximation; factor $3$ before $H$ is calculated by
multiplying $t_{2g}$ spin $3/2$ with Lande factor $2$. Besides, the
influence of $H$ on $D(E)$ has also been taken into account. The average $%
e_{g}$ electron concentration $n\in \lbrack 0,1]$. The DOS, chemical
potential and ground energy of other phases can also be calculated exactly
just as the FM case.

\begin{figure}
\includegraphics[width=0.4\textwidth,trim=5 8 10 23]{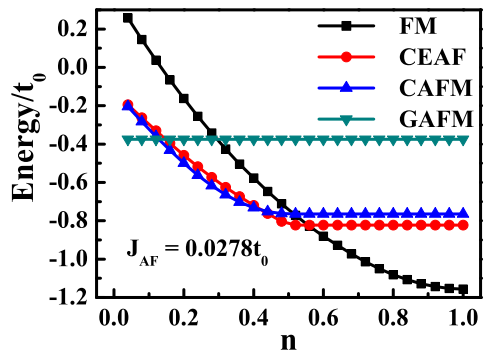}
\caption{\label{Fig.2}(color online) System energy $E$ (ground state
under zero field) as a function of $e_{g}$ electron concentration
$n$, with $J_{AF}=0.0278t_{0}$. Upon increasing $e_{g}$ electrons
$n$, a transition sequence of the ground state: GAFM$\rightarrow
$CAFM$\rightarrow $CE$\rightarrow $FM, is shown. The CE phase as
ground state is possible only in the narrow range around the
half-filling point.}
\end{figure}

However, what should be noted is that none of these phases can be stable
over the whole doping range. In order to determine which phase is the
preferred one at a given concentration, the ground state energy of these
phases, when $J_{AF}$ is set as $0.0278t_{0}$ and $H=0$, is plotted in Fig.%
\ref{Fig.2}. The preferred ground state is the energy minimal state. It
should be GAFM as $n\thicksim 0$ because the interaction is almost pure AFM
super exchange. As $n\thicksim 0.3$, the preferred state is CAFM. The CE
phase can appear only in a narrow regime $n\thicksim 0.5$. When the gain
from kinetic energy suppresses the loss of super exchange energy in the
large $n$ regime, the FM becomes the stable phase. With this calculation,
the phase diagram over the whole concentration regime has be developed. Of
course, the phase diagram of real manganites is more complex than this
simple sketch for two reasons: first, the Hamiltonian Eq.(1) is
oversimplified and secondly the candidate phases chosen here are not
complete but only four phases. Even though, the calculated phase diagram is
quite similar to that of some typical manganites (zero-temperature): e.g. Nd$%
_{1-x}$Sr$_{x}$MnO$_{3}$ (here $n=1-x$) \cite{Kaji}. In fact, it is shown
that the phase diagram of other manganites: e.g. La$_{1-x}$Sr$_{x}$MnO$_{3}$
or Pr$_{1-x}$Ca$_{x}$MnO$_{3}$, can be reproduced roughly by adjusting the
value of $J_{AF}$, whose role will be revisited below. An important truth
revealed here is that no matter what $J_{AF}$, the CE phase can either
appear in the narrow regime $n\thicksim 0.5$ or simply be unstable over the
whole concentration regime.

The above theoretical approach based on Hamiltonian Eq.(1), whose origin can
be found from Ref.\cite{review}, allows us to study the PE in PS regime.
Here, a simple but representative case: the PE between FM phase and CE phase
with $n=0.5$, will be studied. It corresponds to the melting process of a
type of CO state (here it is CE phase) under external magnetic field. Since
for FM metallic phase the lattice distortions are absent and the $e_{g}$
electrons are delocalized \cite{Ahn}, the electron-phonon coupling and
on-site Coulomb repulsion are unimportant. On the other hand, the CE phase
can also be reasonably described as a band insulator by this Hamiltonian at $%
n=0.5$ case \cite{review,Brink}. Therefore, Eq.(1) is suitable to deal with
the PE between FM metallic phase and CE phase, noting that Eq.(1) can be
exactly solved without scale issue.

The PE principle for a PS system can be represented by the equivalence in
chemical potential between the competitive phases, i.e. FM and CE to be
considered. From the above calculation (Fig.\ref{Fig.1}(b)), it is seen that
at $n=0.5$ and under zero field, the chemical potential of pure FM phase is
about $-1.52t_{0}$ and that of CE phase is $-t_{0}$. Therefore, the chemical
potential $\mu $ for the possible PS system would be in the range [$%
-1.52t_{0}$, $-t_{0}$]. The sum of $e_{g}$ electrons in the FM and CE phases
can be calculated:
\begin{equation}
n_{A}=\frac{p_{A}}{N}\int^{E<\mu }D(E)dE  \label{n}
\end{equation}%
here $p_{A}$ is the volume fraction of $A$ (FM/CE) phase. The upper limit of
the integral should meet the equivalent chemical potential condition: $\mu
_{FM}=\mu _{CE}=\mu $. Then the following set of equations yields:
\begin{equation}
\left\{
\begin{array}{c}
p_{FM}+p_{CE}=1 \\
n_{FM}+n_{CE}=0.5%
\end{array}%
\right.  \label{equ}
\end{equation}

\begin{figure}
\includegraphics[width=180pt,height=220pt,trim=15 13 10 23]{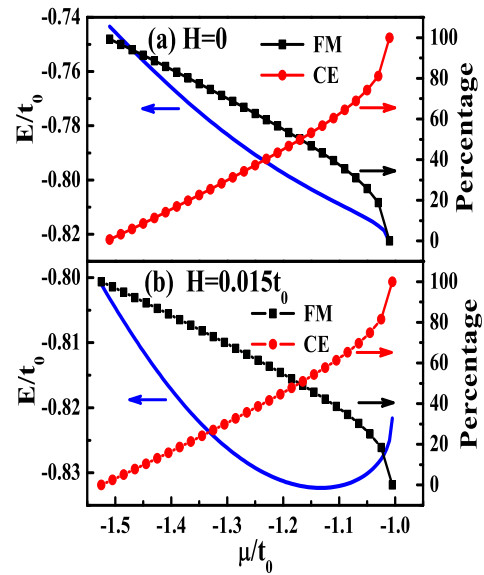}
\caption{\label{Fig.3}(color online) Volume fraction of FM/CE phase
(right axis) and corresponding total energy system $E$ (left
axis) as a function of chemical potential $\protect\mu $. $%
J_{AF}=0.0278t_{0} $. (a)Zero field case. where the ground state is
CE phase without phase separation because the energy decreases
monotonously with chemical potential; (b)under magnetic field
$H=0.015t_{0}$, where a energy minimal appears at $\protect\mu
=-1.13t_{0}$. The coexistence with the fraction of FM and CE phase
in this point is the ground state.}
\end{figure}

By numerical method, Eq.(\ref{equ}) can be solved with the prior assumed $%
\mu $. Given parameters $J_{AF}=0.0278t_{0}$ and $H=0$, the calculated
relative volume fraction of FM/CE phase as a function $\mu $ are shown in
Fig.\ref{Fig.3}(a) (right axis). Then by including the factor $p_{A}$ into
the energy equation, the total system energy: $E=p_{FM}E_{FM}+p_{CE}E_{CE}$
can been calculated, as shown in Fig.\ref{Fig.3}(a) (left axis). It is seen
that both $p_{FM}$ and $E$ are monotonously decreasing functions of $\mu $.
The ground state must be homogenous CE phase because the PS state goes
against energy. In case of nonzero magnetic field (e.g. $H=0.015t_{0}$), the
above calculation is repeated by taking the magnetic field contribution to
DOS and energy into account. The results are plotted in Fig.\ref{Fig.3}(b).
Quite interestingly, a nonzero field results in a minimal in the $E-\mu $
curve (here at $\mu \thicksim -1.13t_{0}$). The PS state associated with
this minimal energy point is more stable with respect to the homogenous FM
or CE phase. The value $-1.13t_{0}$ is the real chemical potential of this
PS ground state which consists of $57\%$ CE phase and $43\%$ FM phase.

\begin{figure}
\includegraphics[width=0.5\textwidth,trim=15 6 10 26]{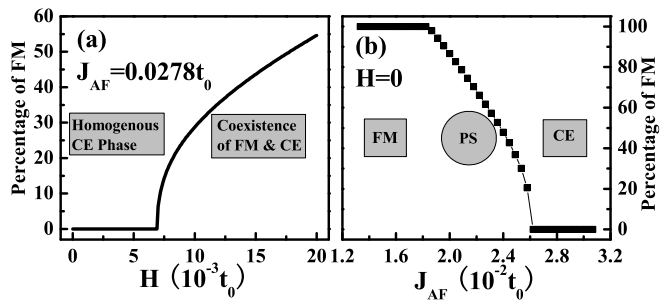}
\caption{\label{Fig.4}(a)Melting process of the CE phase under
magnetic field. When the field is below the threshold $H_{min}$, the
homogenous CE phase is robust. Then phase separation occurs when
$H>H_{min}$. (b) Phase transitions induced by $J_{AF}$ under
zero-field condition: FM phase at weak $J_{AF}$;
coexistence of FM and CE phase at middle $J_{AF}$; CE phase at strong $%
J_{AF} $.}
\end{figure}

By varying the value of $H$, one can repeat the above calculation and then
obtain the relative volume fractions of the two phases as a function of $H$,
as shown in Fig.\ref{Fig.4}(a). When $H$ is low (below the lower threshold $%
H_{min}\thicksim 0.0069t_{0}$), the CE phase is robust against magnetic
field perturbation, indicating the stable and homogenous CE phase as the
ground state. Upon an increasing of $H$ beyond $H_{min}$, part of the CE
phase begins to melt into FM phase, and $p_{CE}$ is deceasing with
increasing $H$ under the equivalent chemical potential condition. In
details, $p_{FM}$ increases rapidly once $H>H_{min}$, and the growth becomes
slower when $H$ is further higher, as shown in Fig.\ref{Fig.4}(a). Note here
that the threshold of percolation for FM phase can be easily surpassed under
a field slightly higher than $H_{min}$. For instance, only a field of $%
H=0.009t_{0}$ is needed to obtain a FM phase of $24.7\%$ (threshold for a
three-dimensional simple cubic bond percolation), beyond which an
insulator-metal transition may be expected. When $H$ is extremely high (not
shown in Fig.\ref{Fig.4}(a)), the equivalent chemical potential condition
can no longer be satisfied, indicating a termination of the PS state and the
ground state will be homogenous FM state.

The above calculation indicates that the lower threshold of magnetic field
to melt the CE phase is $0.0069t_{0}$. Considering that $t_{0}$ in
low-bandwidth manganites is small, e.g. $0.1eV$, the calculated $H_{min}$ is
about $12T$, a value consistent with the experimental data \cite%
{Kuwa,Tomi,Toku,Okim}. Note here that $H_{min}$ is not equal the direct
energy gap between FM and CE phase, which is one order of magnitude larger
than experimental value (shown in Fig.\ref{Fig.2}, about $0.08t_{0}$). It
means that the required magnetic field to destroy the CO insulator is
strongly reduced by PS which can occur once the energy of competitive phases
is close to each other. On the other hand, as identified earlier, parameter $%
J_{AF}$ plays a key role in PS although it is the least intrinsic
interaction in manganites \cite{review}. Other than the above case of
field-induced sequences, abundant phenomena associated with PS in manganites
can be predicted by our model through adjusting $J_{AF}$. For example, again
at $n=0.5$, a coexistence of FM phase and CE phase under zero field is
predicted at $0.019t_{0}<J_{AF}<$ $0.026t_{0}$. When $J_{AF}$ is further
reduced, a homogeneous FM phase as the ground state is possible even under
zero field. The transition between three regimes: homogenous FM phase to PS
state to homogenous CE phase, are identified in Fig.\ref{Fig.4}(b). These
transitions can be mapped to real manganites of wide-band to those of
middle-band and then narrow-band \cite{review}.

It should be mentioned that the only parameter adjustable here is the super
exchange $J_{AF}$, and the energy difference between different phases is
dependent on ratio $J_{AF}/t_{0}$. This is obviously oversimplified,
referring to real manganites materials in which not only the double/super
exchange but also the Jahn-Teller distortion and Coulomb repulsion play
important roles. For instance, the Jahn-Teller distortion in CE phase will
affect the energy band and DOS \cite{Brey,Dong2}. In addition, although the
phase diagram given in Fig.\ref{Fig.2} is quite similar to those for some
manganites, there are still some blemishes. For instance, a prediction of
the correct concentration $n$ corresponding to the A-type AFM observed in
LaMnO$_{3}$ ($n\thicksim 1$) \cite{Wollan} or Nd$_{1-x}$Sr$_{x}$MnO$_{3}$ ($%
x\thicksim 0.55$) \cite{Kaji} can not be given by the present model.
However, if a complete Hamiltonian is employed, the calculation has to be
oversimplified, e.g. limited in a small cluster which is disadvantageous to
deal with PS. Fortunately, Eq.(1) in the present work can describe FM/CE
phase to some satisfactory extent and it is a good starting point to
investigate the PE issue in PS systems against external magnetic field
perturbation. Furthermore, the present approach represents a general roadmap
to investigate the PE issues in manganites: e.g. phase competition other
than FM-CE, of more than two phases, of different $e_{g}$ concentrations,
and effect of other perturbation than magnetic field etc. The key condition
is the equivalence of chemical potential between competitive phases.

In summary, the principle of chemical potential equivalence has been
introduced to investigate the phase equilibrium of half-doped manganites
under external magnetic field. By employing the two orbital model, we have
presented an explicit solution to the phase equilibrium between FM phase and
CE phase. The magnetic field threshold required for melting of the CE phase
has been calculated, consistent with the experimental results. The volume
fractions of the two competitive phases in the phase separation regime as a
function of external magnetic field have been evaluated. In addition, the
super exchange modulated transitions between ferromagnetic, phase separated
and CE states under zero-field, is predicted.

\begin{acknowledgments}
S. Dong thanks G. X. Cao for valuable discussions. This work was supported
by the Natural Science Foundation of China (50332020, 10021001, 10474039)
and National Key Projects for Basic Research of China (2002CB613303,
2004CB619004).
\end{acknowledgments}

\end{document}